# Multimodal Pseudo-CT Synthesis for PET Attenuation Correction using Separate Modality Encoding and Topogram Conditioning

Rory Bell [0009-0002-7778-1231], Artemis Bouzaki[0009-0002-2222-6687], Jiaming Cao[0009-0000-9196-6131], Jasmine Morrison[0009-0005-2564-1327], and Chelsea Sargeant[0000-0002-8737-4835]*

Radiotherapy Related Research, Division of Cancer Sciences, School of Medical Sciences, The University of Manchester, Manchester, United Kingdom

*All authors contributed equally and are presented in alphabetical order.

**Abstract.** We participated in the BIC-MAC Challenge with a multimodal 3D patch-based U-Net for pseudo-CT generation from NAC-PET, MRI, and 2D topograms. By using separate PET and MR encoders, multi-scale feature fusion, and FiLM-based topogram conditioning at bottleneck, we obtain a model that integrates complementary cross-modal information while reducing reliance on precise voxel-wise correspondence between modalities. Our final submission can be found: https://github.com/rrr-uom-projects/BIC-MAC-MICCAI2026

## 1. Introduction

CT is used for attenuation correction in PET imaging but introduces additional ionising radiation. The BIC-MAC Challenge explores the possibility of attenuation correction without CT by generating pseudo-CTs from non-attenuation-corrected PET (NAC-PET), MRI, and 2D topograms using a 75-subject training set. The generated pseudo-CTs are then used for PET reconstruction, making both CT quality and PET correction relevant to model performance.

## 2. Methods

### 2.1 Multimodal architecture

We developed a multimodal 3D patch-based U-Net for pseudo-CT generation from NAC-PET, MRI, and 2D topograms. As image modalities were not perfectly aligned, NAC-PET and two MRIs (in- and out-phase) were processed by separate PET and MRI 3D encoders, rather than early channel-wise fusion. This allowed each modality to learn its own distinct level features before complementary information was combined through multi-scale feature fusion. At each of the four encoder scales, the PET and MR feature maps were channel-wise concatenated, projected back to the single-modality channel width with a 1x1x1 convolution, and reweighted by a squeeze-and-excitation (SE) gate [1]. The SE gate included global average pooling followed by a channel-reduction bottleneck (reduction ratio 4) and a sigmoid, adaptively reweighting the fused representation at each channel and scale. The fused features were propagated through U-Net skip connections and a shared bottleneck.

The 2D topogram was treated differently from the volumetric inputs. As a projection image, it provided global anatomical and body-shape information but had no direct spatial correspondence with individual 3D patches. We used a pretrained ResNet-18 [2] to extract a 256-dimensional feature. The pretrained input convolution was adapted from 3-channel RGB to single-channel topogram input by averaging its weights across the channel dimension. Rather than spatially combining the 2D and 3D inputs, the topogram feature conditioned the PET-MR bottleneck using feature-wise linear modulation (FiLM) [3]. Channel-wise scaling and shifting parameters $\gamma$ and $\beta$ were learned from the topogram features and applied as $b' = b(1+\gamma) + \beta$, where b is the PET-MR bottleneck. The FiLM projection layer was zero-initialised so that $\gamma = \beta = 0$ at the start of training, making the conditioning an identity mapping initially and allowing the influence of the topogram branch to be learned progressively. The conditioned feature was reconstructed through the 3D decoder to produce the final pseudo-CT.

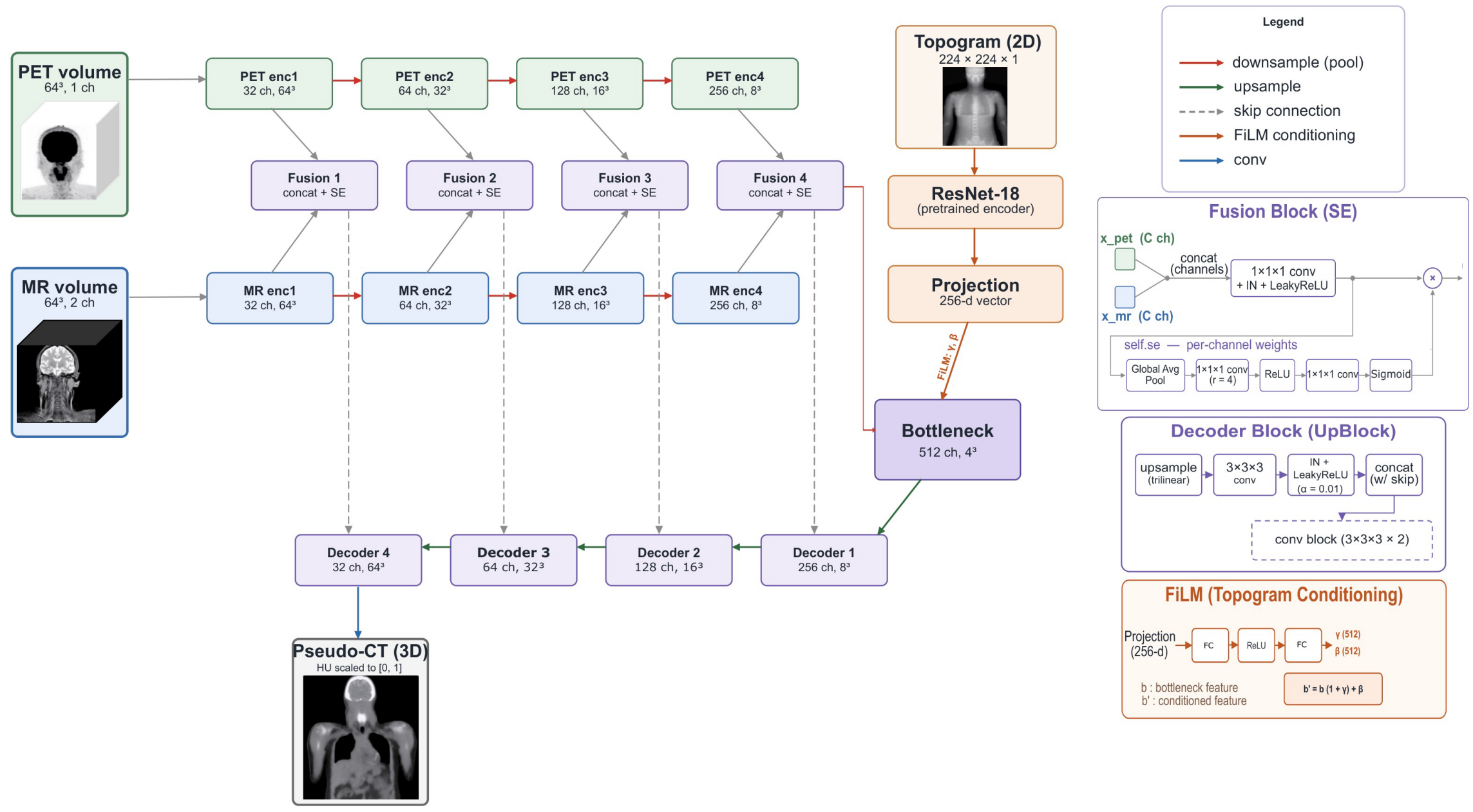

**Fig 1**: Network architecture. Separate 3D encoders process NAC-PET and MRI at four scales; fused features (concatenation + 1×1×1 conv + SE gate) form U-Net skip connections and feed a shared bottleneck. A 2D topogram is encoded with a pretrained ResNet-18 and conditions the bottleneck via zero-initialised FiLM.

## 2.2 Implementation details

The PET and MR encoders, fusion blocks, bottleneck, and decoder used a base channel width of 32, giving encoder/decoder feature widths of 32-64-128-256 across the four scales and a 512-channel bottleneck. Each convolutional block used two 3x3x3 convolutions with instance normalisation and LeakyReLU (negative slope 0.01). Downsampling used 2x max pooling, while the decoder upsampled with resize convolutions (trilinear upsampling followed by convolution) rather than transposed convolutions to avoid checkerboard artifacts [4]. Each upsampled feature map was then concatenated with the corresponding fused encoder features before a further convolutional block. The final 1x1x1 convolution was followed by a sigmoid, consistent with all inputs and the target CT being normalised to [0,1].

All volumetric inputs were intensity-normalised before patch extraction. NAC-PET and the two MRIs were independently scaled from the 0.5–99.5 intensity percentiles to [0,1], while CT values were clipped to [−1000,2000] HU and linearly mapped to [0,1]. The topogram was resized to 224×224. Of the 75 subjects with ground truth CT, 71 were used for training and 4 for internal validation. During training, 64×64×64 patches were sampled using positive-only foreground sampling from the provided prediction mask, with a batch size of 1, for 250 epochs with AdamW optimiser (learning rate of 1e-4, weight decay 1e-5, cosine annealing). The model was trained using an L1 loss evaluated within the prediction mask. Automatic mixed precision was used with gradient clipping at norm 1.0. Training was performed on a single NVIDIA RTX A6000 48GB for approximately 9 hours, using PyTorch 2.11.0 and MONAI 1.6.0. Full-volume inference used MONAI sliding-window inference with 64×64×64 patches, 50% overlap, and Gaussian blending of overlapping predictions, requiring approximately 150s per subject.

### 2.3 Augmentation and architectural ablations

We evaluated this configuration against several alternatives: affine and intensity data augmentation (jointly applied to PET and MR to preserve spatial correspondence), attention-based fusion, alternative convolutional blocks, weighted loss functions, a learned rigid registration network for MR–CT spatial correspondence, simultaneous training with registered and non-registered images, along with region-specific MRI masking. None of these improved CT or PET reconstruction metrics over the configuration described above on our internal validation subject, and the best-performing augmented configuration we tested was consistently worse than the unaugmented model on whole-body CT MAE and organ bias. Our final submission therefore uses the model described above, trained without data augmentation.

## 3. Results

The quantitative results of the final submission model on the validation set are shown in Table 1. The model achieved a CT mu-map MAE of 0.005454 and a whole-body PET SUV MAE of 0.0379. For the region-specific PET metrics, the mean organ bias was 2.45%, while the brain outlier score was 0.024. Figure 2 shows pseudo-CT outputs for four subjects from the internal validation set. The provided prediction masks were applied for visualisation to remove regions outside the body.

| Metric | [CT] Mu-map MAE | [PET] Whole-body SUV MAE | [PET] Organ Bias % | [PET] Brain Outlier Score |
|---|---|---|---|---|
| **Score** | 0.005454 | 0.0379 | 2.45 | 0.024 |

**Table 1.** CT and PET metrics score of submission model on the validation set.

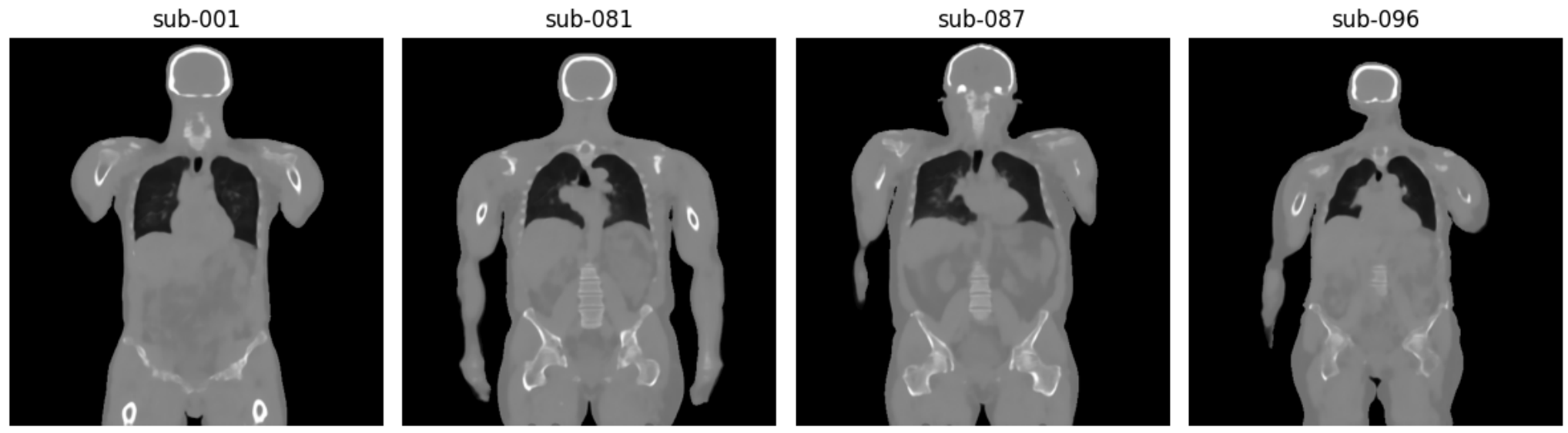

**Fig 2.** Coronal views of pseudo-CTs generated for four internal validation subjects.

## 4. Conclusion

We developed a multimodal approach for pseudo-CT generation from NAC-PET, MRI, and 2D topograms. The proposed architecture combines modality-specific volumetric feature extraction with topogram-based conditioning to integrate complementary cross-modality information. The validation results demonstrate promising performance for downstream PET attenuation correction.